\documentclass[12pt,preprint]{aastex}

\def\kms{km\,s$^{-1}$}

\def\gtrsim{\mathrel{\hbox{\rlap{\hbox{\lower4pt\hbox{$\sim$}}}\hbox{$>$}}}}
\def\lesssim{\mathrel{\hbox{\rlap{\hbox{\lower4pt\hbox{$\sim$}}}\hbox{$<$}}}}

\shorttitle{Progenitor of SN~2009hd}
\shortauthors{Elias-Rosa et al.}

\begin{document}


\title{The Massive Progenitor of the Possible Type II-Linear Supernova 2009hd in Messier 66}

\author{Nancy Elias-Rosa\altaffilmark{1,2,3},
  Schuyler D.~Van Dyk\altaffilmark{1},
  Weidong Li\altaffilmark{2},
  Jeffrey M.~Silverman\altaffilmark{2},
  Ryan J. Foley\altaffilmark{4,5},
  Mohan Ganeshalingam\altaffilmark{2},
  Jon C.~Mauerhan\altaffilmark{1}, 
  Erkki Kankare\altaffilmark{6,7},
  Saurabh Jha\altaffilmark{8},
  Alexei V. Filippenko\altaffilmark{2},
   John E. Beckman\altaffilmark{9,10},
   Edo Berger\altaffilmark{4},  
  Jean-Charles Cuillandre\altaffilmark{11}, and
  Nathan Smith\altaffilmark{2,12}.
  }

\altaffiltext{1}{Spitzer Science Center, California Institute of Technology, 1200 E. California Blvd., Pasadena, CA 91125, USA; email nelias@ice.csic.es.}
\altaffiltext{2}{Department of Astronomy, University of California, Berkeley, CA 94720-3411, USA.}
\altaffiltext{3}{Institut de Ci\`encies de lÕEspai (IEEC-CSIC), Facultat de Ci\`encies, Campus UAB, 08193 Bellaterra, Spain}
\altaffiltext{4}{Harvard/Smithsonian Center for Astrophysics, 60 Garden Street, Cambridge, MA 02138, USA.}
\altaffiltext{5}{Clay Fellow.}
\altaffiltext{6}{Tuorla Observatory, Department of Physics and Astronomy, University of Turku, V\"{a}is\"{a}l\"{a}ntie 20, 21500 Piikki\"{o}, Finland.} 
\altaffiltext{7}{Nordic Optical Telescope, Apartado 474, E-38700 Santa Cruz de La Palma, Spain.}
\altaffiltext{8}{Department of Physics and Astronomy, Rutgers, The State University of New Jersey, 136 Frelinghuysen Road, Piscataway, NJ 08854, USA.}
\altaffiltext{9}{Instituto de Astrofisica de Canarias, E-38200 La Laguna, Tenerife, Spain.}
\altaffiltext{10}{C.S.I.C., Madrid, Spain.}
\altaffiltext{11}{Canada-France-Hawaii Telescope Corporation, 65-1238 Mamalahoa Hwy, Kamuela, HI 96743, USA.}
\altaffiltext{12}{Steward Observatory, University of Arizona, Tucson, AZ 85721-0065, USA.}

\begin{abstract}
We present early- and late-time photometric and spectroscopic
observations of supernova (SN)~2009hd in the nearby spiral galaxy
NGC~3627 (M66). This SN is one of the closest to us in recent years
and provides an uncommon opportunity to observe and study the nature of
supernovae. However, the object was heavily obscured by dust, rendering it unusually
faint in the optical, given its proximity. We find that the observed
properties of SN~2009hd support its classification as a possible Type II-Linear
SN (SN~II-L), a relatively rare subclass of core-collapse
supernovae. High-precision relative astrometry has been employed to attempt to
identify a SN progenitor candidate, based on a pixel-by-pixel
comparison between {\sl Hubble Space Telescope\/} ({\sl HST}) F555W
and F814W images of the SN site prior to explosion and at late
times. A progenitor candidate is identified in the F814W images only; this object is undetected in 
F555W.
Significant uncertainty exists in the astrometry, such that we cannot definitively identify
this object as the SN progenitor.
Via insertion of artificial stars into the pre-SN {\sl HST\/} images, we are able to constrain
the progenitor's properties to those of a possible supergiant, with intrinsic absolute magnitude $M_{F555W}^0 \gtrsim -7.6$ mag and 
intrinsic color $(V-I)^0 \gtrsim 0.99$ mag. The magnitude and color limits are consistent with a
luminous red supergiant; however, they also allow for the possibility that 
the star could have been more yellow than red.
From a comparison with
theoretical massive-star evolutionary tracks which include rotation and pulsationally enhanced 
mass loss, we can place a conservative upper limit on the
initial mass for the progenitor of
$M_{\rm ini} \lesssim 20\ {\rm M}_{\sun}$. If the actual mass of the progenitor is near the upper range allowed by our derived mass limit, then it would be
consistent with that for the identified progenitors of the SN II-L 2009kr and the high-luminosity SN II-Plateau (II-P) 2008cn.
The progenitors of these three SNe may possibly bridge the gap
between lower-mass red supergiants that explode as SNe II-P
and luminous blue variables, or more extreme red supergiants,
from which the more exotic SNe~II-narrow may arise. Very late-time imaging of the SN~2009hd site may provide us with more clues regarding the true nature of its progenitor.
\end{abstract}


\section{Introduction}\label{introduction}

The Type II-Linear supernovae (SNe~II-L) are among the least common
and, therefore, most poorly studied subclasses of core-collapse
supernovae (CC-SNe). They represent 6.4--10\% of all CC-SNe
\citep{smith10,li10}. The spectra of SNe~II-L are similar to those of
the much more common SNe~II-Plateau (II-P), but SNe~II-L are
distinguished by the shapes of their light curves. While light curves
of SNe~II-P exhibit a relatively constant plateau for $\sim$100~d
after an initial peak in the first few days, the light curves of
SNe~II-L instead show a linear decline commencing just after the
initial peak (e.g., \citealt{barbon79}). The very short, or
nonexistent, plateau phase for SNe~II-L may well result from a
low-mass H envelope at the time of explosion (between 1 and
5~M$_{\sun}$; \citealt{swartz91}). The envelope can be further reduced
through pre-SN mass loss, and if so, we would infer that the mass
of the progenitor star is higher than that expected for SNe~II-P,
$8.5^{+1.0}_{-1.5} \lesssim M_{\rm ini} \lesssim 16.5 \pm
1.5$~M$_\odot$ \citep{smartt09}. The direct identification of the
progenitor of the SN~II-L~2009kr \citep{eliasrosa10} provides evidence
that this indeed may be the case (note that \citealt{fraser09} suggest that this SN is a spectrally peculiar SN II-P). Clearly, the identification of the
progenitors of other SNe~II-L is necessary to establish a trend, if
one exists.

\begin{figure*}
\centering
\includegraphics[height=4truein,width=4.5truein]{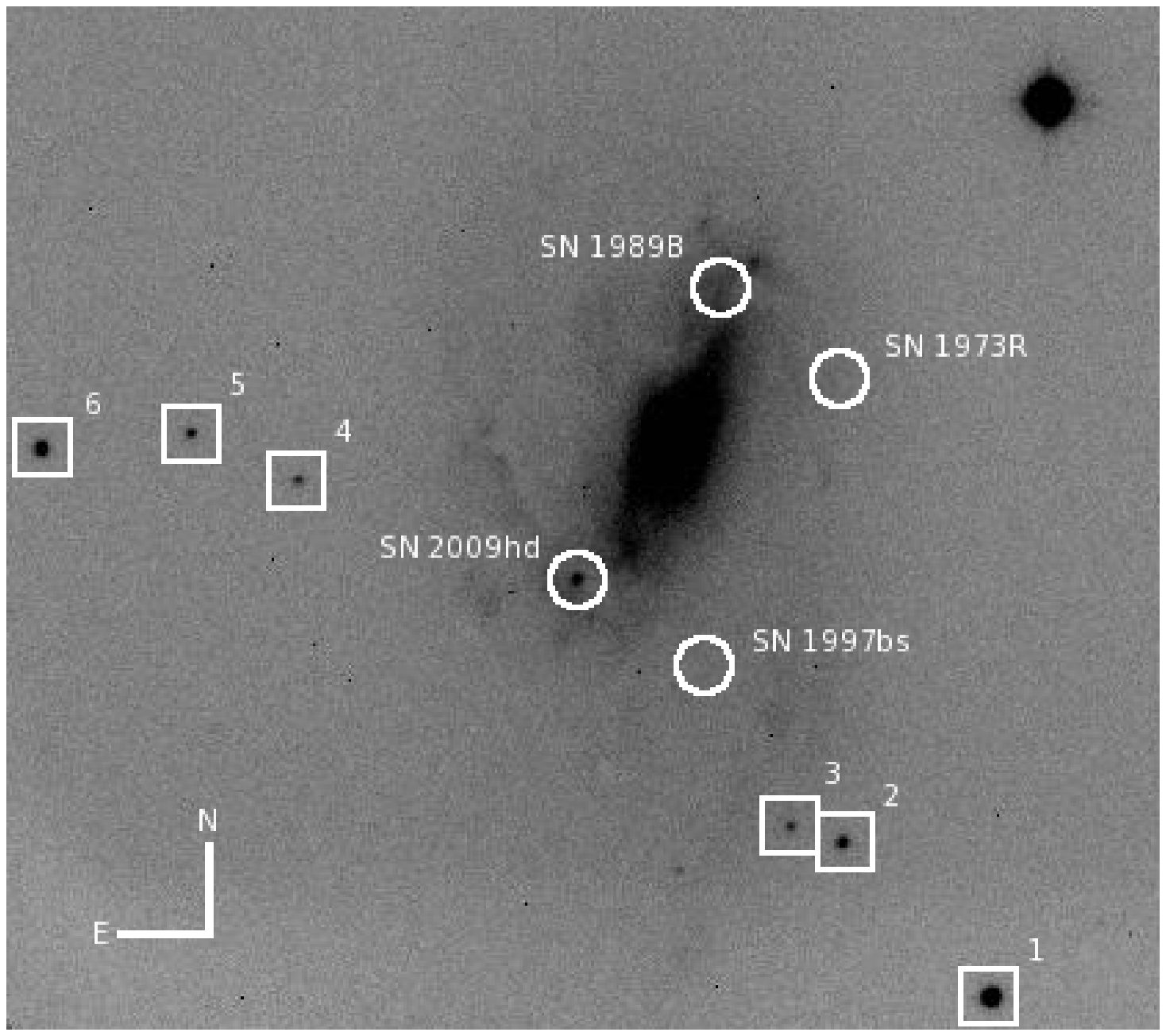}
\caption{$I$-band image of SN~2009hd in NGC~3627 (M66) obtained with
  the 1.3~m SMARTS telescope+ANDICAM at CTIO on 2009 July 10 (field of
  view $\sim 6\arcmin \times 6\arcmin$). The sites of SNe~1973R, 1989B,
  and 1997bs, as well as the local photometric sequence stars, are
  indicated.}
\label{fig_seq}
\end{figure*}

In this paper, we present the case of SN~2009hd ($\alpha = 11^{\rm
  h} 20^{\rm m} 16{\fs}90$, $\delta = +12\arcdeg 58\arcmin
47{\farcs}1$; J2000.0) in NGC 3627 (Messier 66, M66). This SN was discovered by
\cite{monard09} on 2009 July 2.69 (UT dates are used throughout) and
was spectroscopically classified as a reddened SN~II
\citep{kasliwal-et-al-09} with slow evolution \citep{kasliwal09}, and,
subsequently, as a SN~II-P \citep{berger09}. \cite{maoz09} first
identified a possible progenitor star in high spatial resolution
archival {\sl Hubble Space Telescope\/} ({\sl HST\/}) images obtained
between 1997 and 1998. Here we show that both the photometric and
spectroscopic observations suggest that this object is, in fact, a
SN~II-L. We also present a substantially improved identification of the
progenitor, based on a pixel-by-pixel comparison between the pre-SN
{\sl HST\/} data and late-time {\sl HST\/} images of the SN. We note
that M66 has been host to two other known SNe, the SN~II~1973R
\citep{ciatti77} and the Type Ia~SN~1989B \citep{wells94}, as well as
the ``SN impostor'' SN 1997bs \citep{vandyk00}; see Figure \ref{fig_seq}.


\section{The Nature of SN~2009hd}\label{specph}

\subsection{Photometry}\label{specph_ph}

Optical ($BVRI$) images of SN~2009hd were obtained with the 1.3~m
SMARTS+ANDICAM at Cerro Tololo Inter-American Observatory (CTIO), the
Palomar 1.5~m telescope, the 2.0~m Liverpool Telescope and the 2.6~m
Nordic Optical Telescope (NOT) at the Roque de los Muchachos
Observatory, the 6.5~m Baade Magellan telescope, and the 10~m Keck-II
telescope. The photometric observations were processed using
IRAF\footnote[13]{IRAF (Image Reduction and Analysis Facility) is
  distributed by the National Optical Astronomy Observatories, which
  are operated by the Association of Universities for Research in
  Astronomy, Inc. (AURA), under cooperative agreement with the
  National Science Foundation (NSF).} routines with the standard
recipe for CCD images (trimming, overscan, bias, and flat-field
corrections). Due to the location of SN~2009hd along a bright spiral
arm of M66, contamination of the photometry by the host-galaxy light
was a serious problem. We therefore used the template-subtraction
technique to remove this background and hence to measure more
accurately the SN magnitudes. The template images of M66 were obtained
with the Kitt Peak National Observatory 2.1~m telescope, under
$1\arcsec$ seeing conditions, about seven years prior to the SN
discovery \citep{kennicutt03}. The instrumental magnitudes of the SN
and the reference stars in the SN field were measured in the
subtracted images using the point-spread function (PSF) fitting
technique with the DAOPHOT package \citep{stetson87} within IRAF.

In order to calibrate the instrumental magnitudes to the standard
photometric system, we used the magnitudes and colors of the local
sequence stars in the SN field (Figure \ref{fig_seq} and Table
\ref{table_seq}). These values were measured from $BVRI$ images of NGC
3627 obtained under photometric conditions on 1999 December 17 with the
1.5~m telescope at Palomar Observatory \citep[see][]{vandyk00}.
 
\begin{figure*}
\centering
\includegraphics[height=5truein,width=3.8truein,angle=-90]{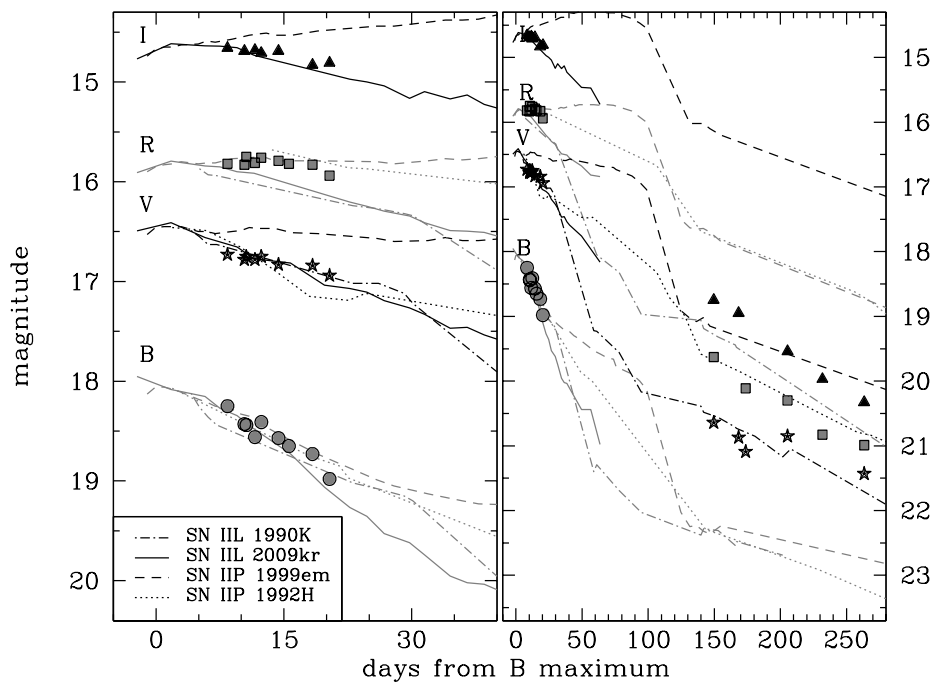}
\caption{{\it Left:} Early-time ( $\lesssim$ 40~d after $B$-band maximum)
  optical light curves of SN~2009hd, together with those of the
  SNe~II-L~1990K ({\it dot-dashed lines}) and 2009kr ({\it solid
    lines}), and of the SNe~II-P~1999em ({\it dashed lines}) and 1992H
  ({\it dotted lines}). {\it Right:} Optical light curves of
  SN~2009hd, including late-time data. The comparison light curves are
  adjusted in time and magnitude to match those of SN 2009hd. }
\label{fig_lcurve}
\end{figure*}

SN~2009hd was observed during the first $\sim$20~d after discovery, but
it subsequently became too close to the Sun in the sky. It was
observed again five months later (Table~\ref{table_ph}). From the discovery dates given by
\citet{monard09} and \citet{yamaoka09}, we constrained the explosion
date to be 2009 June $19 \pm 3$, or JD 2,455,002 $\pm 3$. From the
light curves of a sample of SNe~II-L, it takes $\sim 9$~d from
explosion to $B$ maximum \citep{li10}. The $BVRI$ light curves for
SN~2009hd are shown in Figure \ref{fig_lcurve}, relative to $B$
maximum (2009 June $28 \pm 3$, or JD 2,455,011 $\pm 3$). For
comparison, we also show the light curves of the SNe~II-L~1990K
\citep{cappellaro95} and 2009kr \citep{eliasrosa10}, and the
SNe~II-P~1999em \citep{hamuy01,leonard02,elmhamdi03} and 1992H
\citep{clocchiatti96}. From the light-curve comparison, it appears
that SN~2009hd does not follow a standard SN~II-P evolution
and does not behave like
any SN~IIb \citep[e.g., SN~1993J;][not shown in the 
figure]{barbon95}. Instead, it more closely resembles the SN~II-L
post-maximum decline. During early epochs, the light curves show a steep,
almost linear, behavior. 
A linear fit to
the data for
the first 12~d results in decline rates $\beta^{B}_{12} = 0.62 \pm 0.05$, $\beta^{V}_{12} = 0.18 \pm
0.03$, $\beta^{R}_{12} = 0.14 \pm 0.03$, and $\beta^{I}_{12} = 0.17
\pm 0.02$ mag (12~d)$^{-1}$ in each band. We note the presence of a ``bump" in the
$BVR$ light curves at $\sim 10$--20~d, which seems to follow the
``shoulder'' exhibited by SN~1990K \citep{cappellaro95}. This
light-curve feature appears in the case of several SNe~II-L
\citep{patat94}. After $\sim 100$~d following $B$ maximum, the
luminosity decline settles down at a significantly lower rate. This
behavior is consistent with complete trapping of the energy release
from the radioactive decay of $^{56}$Co to $^{56}$Fe. In the phase
range 140--260~d the decline rates are $\gamma^{V} = 0.57 \pm 0.08$,
$\gamma^{R} = 1.19 \pm 0.10$, and $\gamma^{I} = 1.44 \pm 0.12$ mag
(100~d)$^{-1}$, which are similar to those found for SN~1990K
\citep{cappellaro95}.

\begin{figure*}
\centering
\includegraphics[height=5truein,width=3.8truein,angle=-90]{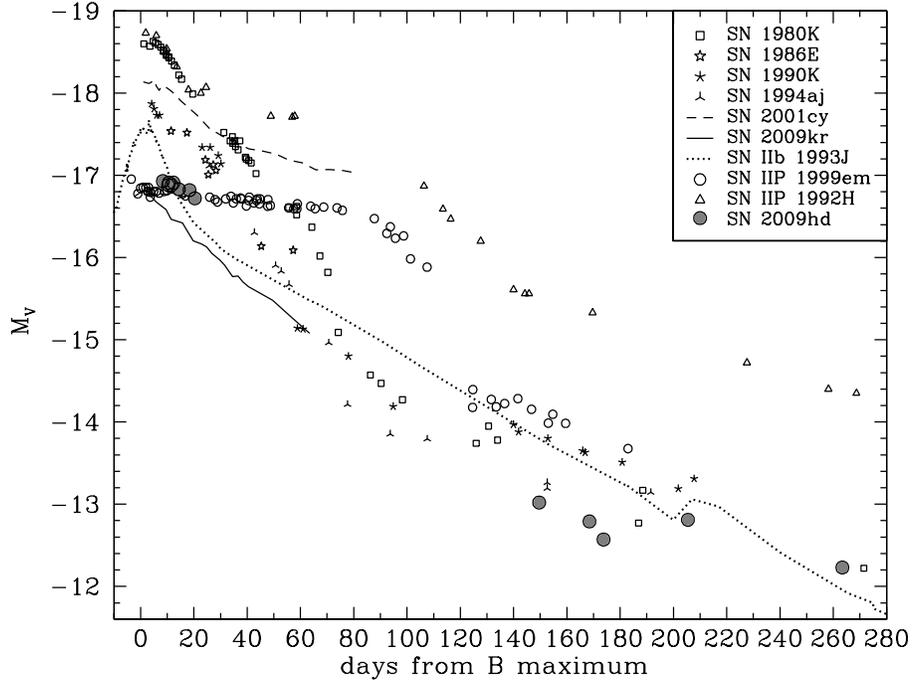}
\caption{Absolute $V$ light curve of SN~2009hd ({\it filled circles}) along with those of the SNe~II-L 1980K ({\it empty
    squares}), 1986E ({\it empty stars}), 1990K ({\it 5-pointed
    stars}), 2001cy ({\it dashed lines}), and 2009kr ({\it solid
    lines}); the peculiar SN II-L 1994aj ({\it 3-pointed stars}); the
  SN IIb 1993J ({\it dotted lines}); and the SNe~II-P~1999em ({\it
    empty circles}) and 1992H ({\it empty triangles}). Distances and
  extinction estimates for all SNe, except SN~2009hd, have been adopted
  from the literature. The magnitudes of SN~2009hd have been corrected for
  the assumed reddening, $E(B-V) = 1.23$ mag, and the adopted
  distance modulus, $\mu = 29.86 \pm 0.08$ mag \citep{freedman01}.} 
\label{fig_absmag}
\end{figure*}

Figure \ref{fig_absmag} shows a comparison between the evolution of 
the absolute $V$ magnitude, $M_V$, of SN~2009hd, compared with the SNe~II-L~1980K
\citep{barbon82,buta82,tsvetkov83}, 1986E \citep{cappellaro90}, 1990K
\citep{cappellaro95}, 1994aj \citep{benetti98}, 2001cy (unpublished
KAIT data), and 2009kr \citep{eliasrosa10}, with the SN~IIb~1993J
\citep{barbon95}, and with the well-studied SNe~II-P~1992H
\citep{clocchiatti96} and 1999em
\citep{hamuy01,leonard02,elmhamdi03}.
The comparison SNe have been corrected
for extinction using published estimates and assuming the
\citet{cardelli89} extinction law. For SN~2009hd we have adopted a
total extinction of $A_V = 3.80 \pm 0.14$ mag (see \S
\ref{specph_extinc}) and a distance to M66 of 9.4~Mpc (derived from
the extinction-corrected Cepheid distance;
\citealt{freedman01}). Again, it is possible to see the remarkable
similarity between the SNe~II-L and SN~2009hd, primarily with
SNe~1990K and 2009kr, implying that this SN also could be a member of
this rare subclass of CC-SNe. 

\begin{figure*}
\centering
\includegraphics[height=5truein,width=3.8truein,angle=-90]{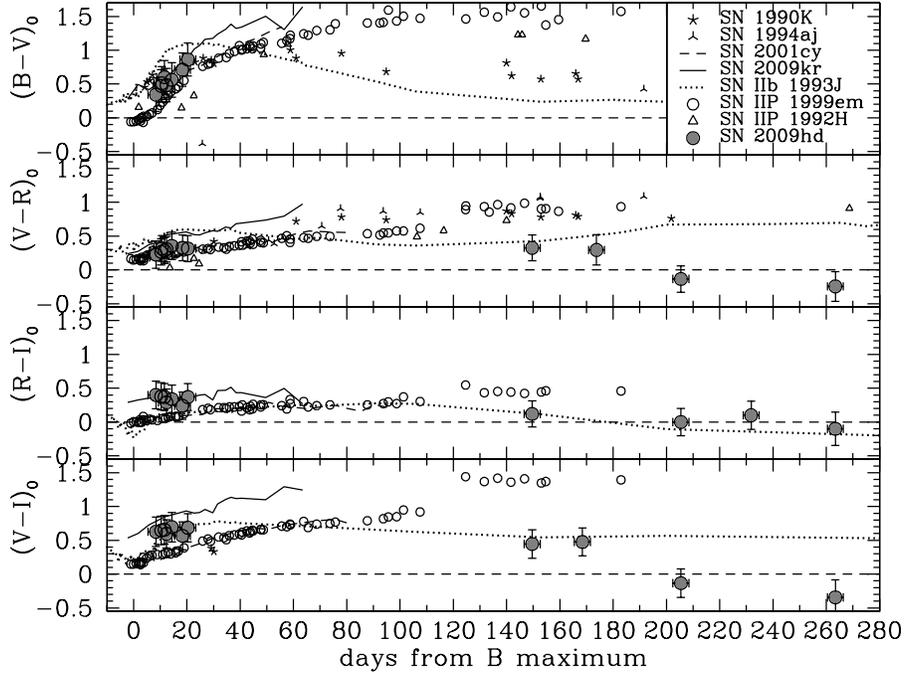}
\caption{Intrinsic $(B-V)_0$, $(V-R)_0$, $(R-I)_0$, and $(V-I)_0$
  color curves of SN~2009hd, compared with the intrinsic color
  evolution of the SNe~II-L 1990K ({\it 5-pointed stars}), 2001cy
  ({\it dashed lines}), and 2009kr ({\it solid lines}); the peculiar
  SN II-L 1994aj ({\it 3-pointed stars}); the SN IIb 1993J ({\it
    dotted lines}); and the SNe~II-P~1999em ({\it empty circles}) and
  1992H ({\it empty triangles}). Distances and extinction estimates
  for all SNe, except SN~2009hd, have been adopted from the
  literature. The color of SN~2009hd has been corrected for the
  assumed reddening, $E(B-V) = 1.23$ mag, and the adopted
  distance modulus, $\mu = 29.86 \pm 0.08$ mag \citep{freedman01}.  }
\label{fig_color}
\end{figure*}

Furthermore, the absolute $(B-V)_0$, $(V-R)_0$,
$(R-I)_0$, and $(V-I)_0$ color evolution of SN~2009hd are typical of a
SN~II-L (see Figure \ref{fig_color}).
The color curves show a rapid evolution from blue to red in the
earlier epochs, except for the $(R-I)_0$ color curves, which initially
exhibit a color decrease for 2--3~d, probably due to the $R$-band
``bump."  At late times, however, the color becomes significantly more
blue than that of other SNe~II. Only the $(R-I)_0$ color curves for
some SNe~IIb, such as SNe 1993J or 2008ax (not shown in the figure;
\citealt{taubenberger11}), have a similar behavior at such
epochs. This apparent bluer color SN 2009hd is possibly 
caused by light from neighboring young stars contaminating our late-time
measurements when the SN is faint.

We therefore conclude that SN~2009hd displays the photometric behavior of a
SN~II-L. We caution, however, that a late plateau may have evaded detection, as a result of 
the absence of data between 30~d and 100~d.
This could imply that SN~2009hd is actually a hybrid event with
characteristics of SNe~II-L and SNe~II-P, such as SN~1992H
\citep{clocchiatti96}.

\subsection{Spectroscopy}\label{specph_spec}

Medium-resolution optical spectra of SN~2009hd were obtained on 2009
July 9 with the Magellan/Clay 6.5~m telescope using the Magellan
Echellete \citep[MagE;][]{marshall08} spectrograph, and on 2010
February 15 with the 10-m Keck-II telescope using the Deep Imaging
Multi-Object Spectrograph \citep[DEIMOS;][]{faber03}.

Standard CCD processing and spectrum extraction of the MagE data were
accomplished with IRAF routines. The data were extracted using the
optimal algorithm of \citet{horne86}. Low-order polynomial fits to
calibration-lamp spectra were used to establish the wavelength
scale. Small adjustments derived from night-sky lines in the object
frames were applied. For the MagE spectra, the sky was subtracted from
the images using the method described by \citet{kelson03}. We employed
IRAF and our own routines to flux calibrate the data and remove
telluric lines using the well-exposed continua of spectrophotometric
standards \citep{wade88, foley03, foley09}.  CCD processing of the
DEIMOS spectra was performed with a modified version of the DEEP
pipeline \citep[e.g.,][]{weiner05}. This produced rectified,
sky-subtracted two-dimensional spectra from which one-dimensional
spectra were then extracted optimally \citep{horne86}. The absolute
flux calibration of the spectra was checked against the photometry
and, when necessary, the spectra were rescaled. After that, the
typical deviation of the spectra from photometry is $< 10$\% in all
bands. We note, however, that the early-time spectrum was observed
$\sim 45\arcdeg$ from the parallactic angle \citep{fil82} at
airmass 2.1, and the flux of the late-time spectrum might be
slightly too low because of the presence of a dead column in the CCD
next to the trace of the SN; consequently, the shape of the continuum
in both spectra could be affected.

A near-infrared (NIR) spectrum was obtained with the Palomar Hale 5~m
telescope using TripleSpec \citep{herter08} on 2009 July 15.  After standard CCD
processing and before extraction, the contribution of the night-sky
lines was removed from one two-dimensional NIR spectral image,
subtracting another two-dimensional spectral image with the SN placed
in a different position along the slit.  In the NIR, the wavelength
calibration was performed using the OH sky lines, and a telluric
absorption correction was derived from observations of the A0V star 
HR~6944 and applied to the SN data using the IDL package 
{\it xtellcor} \citep{vacca03}.

\begin{figure*}
\centering
\includegraphics[height=5truein,width=3.8truein,angle=-90]{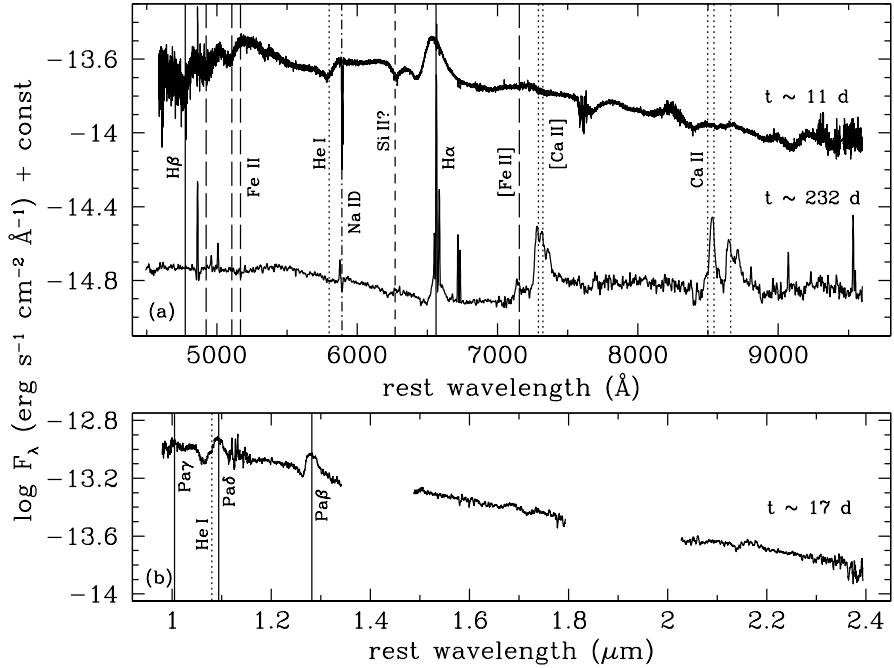}
\caption{(a) Optical spectral evolution of SN~2009hd. The late-time
  spectrum appears to have been 
  contaminated by light from stars near
  the SN. (b) A near-infrared spectrum obtained using TripleSpec with
  the Palomar 5~m Hale telescope on 2009 July 15. Spectra in both
  panels have been corrected for their host-galaxy recession
  velocities, but not for extinction. Ages are relative to $B$ maximum
  light. The locations of the most prominent spectral features are
  marked.}
\label{fig_spectrum}
\end{figure*}

Figure \ref{fig_spectrum} shows the final optical and NIR spectra of
SN~2009hd. The most notable feature of the early optical spectrum
(Figure \ref{fig_spectrum}, top panel) is the very red continuum, 
due to the appreciable extinction of this SN (see \S
\ref{specph_extinc}). The spectrum also exhibits relatively weak
spectral features, mainly due to metals, and a broad H$\alpha$
emission component. The absence of a standard P-Cygni absorption
component of H$\alpha$ has also been observed in some other SNe~II-L,
such as SNe~1979C and 1980K, and this has been postulated as a
characteristic that distinguishes spectroscopically the SNe~II-L from
the SNe~II-P \citep{schlegel96,filippenko97}.

An intriguing feature is present on the blue side of H$\alpha$, at
about 6300~\AA. Similar absorption is also visible in early-time 
spectra of SNe 1999em \citep{baron00,dessart05} and 2005cs
\citep{pastorello06}. This feature has been interpreted either as a
high-velocity H\,{\sc i} line, due possibly to an outer H-rich shell,
or as Si\,{\sc ii} $\lambda$6355 absorption. Unfortunately, the lack
of spectra at early times does not allow us to definitively determine
which of these explanations is correct.  However, since 
the expansion velocities of the ejecta ($\sim$ 6400 and 3600 \kms, respectively) for the 
H$\alpha$ and the putative Si\,{\sc ii} features for SN 2009hd are similar
as those for SN~2005cs at approximately the same epoch 
(see Figure 6 of \citealt{pastorello06}), we
surmise that the absorption near 6300~\AA\ could be more likely due to
Si\,{\sc ii}, rather than to high-velocity hydrogen.

In the late-time spectrum, the continuum has become much bluer than in
the early-time spectrum,
quite probably as a result of contamination by
nearby, young stars, which would have been included within the
spectrograph slit, given the  
seeing conditions. The
spectrum shows weaker H$\alpha$, [Fe\,{\sc ii}] $\lambda 7155$,
Ca\,{\sc ii} $\lambda\lambda7291$, 7324, and [Ca\,{\sc ii}]
$\lambda\lambda8498$, 8542, 8662 emission lines (see Figure
\ref{fig_spectrum}). Since the SN is very close to an H\,{\sc ii}
region, which was also included in the slit, narrow emission lines of
H$\alpha$, H$\beta$, [O\,{\sc iii}] $\lambda5007$, [N\,{\sc ii}]
$\lambda\lambda6548$, 6584, and [S\,{\sc ii}] $\lambda\lambda6717$,
6731 appear superposed on the spectrum.

Finally, the spectrum in the bottom panel of Figure \ref{fig_spectrum}
may well be the first-ever NIR spectrum obtained for a SN~II-L. We
identified lines by comparing the spectrum to that of SNe~II-P
such as SN 2002hh \citep{pozzo06}. The SN 2009hd spectrum is dominated
by lines of the hydrogen Paschen series. The wavelength regions
$\sim$1.35--1.50 $\mu$m and $\sim$1.80--2.00 $\mu$m were compromised,
however, by strong telluric absorption. The relatively featureless
nature of this spectrum is reminiscent of the early-time NIR spectra
of SN~1993J \citep{matthews02}.

\begin{figure*}
\centering
\includegraphics[height=5truein,width=3.8truein,angle=-90]{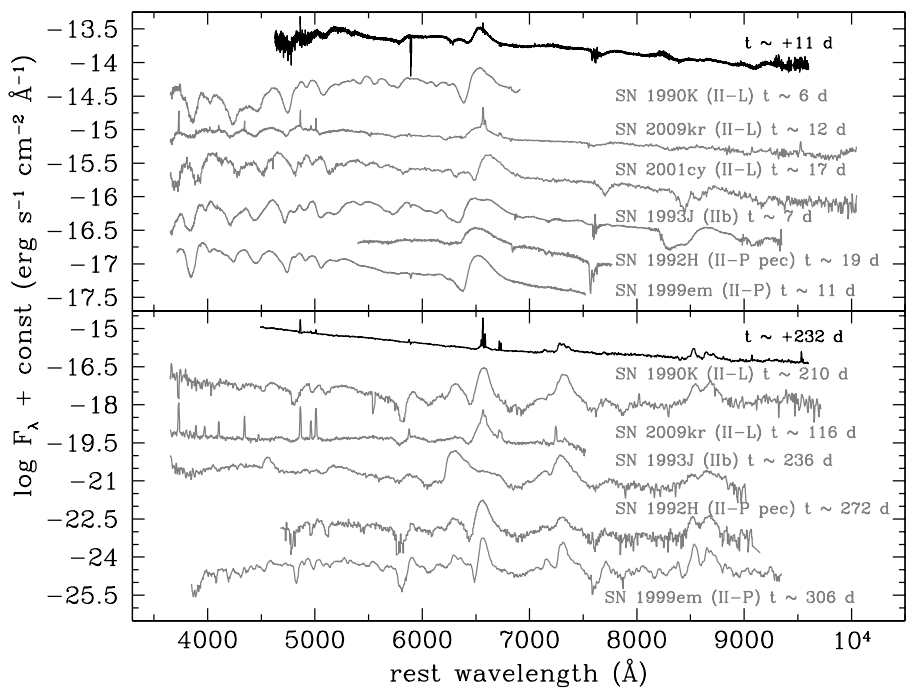}
\caption{Early and late-time optical spectra of SN~2009hd (the top
  spectrum in each panel, shown in bold), along with comparison
  spectra of the SNe~II-L 1990K and 2009kr, SN~IIb~1993J,
  SNe~II-P~1992H and 1999em, and SN~II~2001cy. All spectra have been
  corrected for their host-galaxy recession velocities and for
  extinction (see text). Ages are relative to $B$ maximum light.}
\label{fig_conf}
\end{figure*}

The SN~2009hd optical spectra are shown in Figure~\ref{fig_conf},
relative to the spectra at similar epochs of the SNe~II-L 1990K
\citep{cappellaro95} and 2009kr \citep{eliasrosa10}, as well as of the
SN~IIb~1993J \citep{barbon95}, the SNe~II-P~1992H
\citep{clocchiatti96} and 1999em \citep{leonard02}, and the
SN~II~2001cy (unpublished data). All of the spectra are shown relative
to $B$ maximum; they have been corrected for extinction using
published estimates and deredshifted using values from
NED\footnote[14]{NED, NASA/IPAC Extragalactic Database;
  http://nedwww.ipac.caltech.edu/.}. At all times, SN~2009hd clearly
differs from SNe~1993J, 1992H, and 1999em, and more closely resembles
SNe~2001cy and 2009kr, which also exhibit relatively weak lines (such
as the weak H$\alpha$ absorption trough). SN~1990K, also a SN~II-L,
shows overall stronger features than SN~2009hd. At late times, the
features present for SN~2009hd are similar to those of normal
SNe~II-P, except that they are weaker or nearly absent. The fact that
the continuum is dominated by relatively narrow Ca\,{\sc ii} and
[Ca\,{\sc ii}] emission may imply the obscuration of the SN progenitor
by dust, as well as the destruction of dust in the circumstellar
medium (CSM); see for example the case of SN~2008S
\citep{prieto08,smith09}. We also compared SN~2009hd with the
prototypical Type II-narrow (IIn) SN1988Z (not shown in the figure); they appear
similar blueward of
H$\alpha$, but quite different on the redward side. Thus, once again we emphasize the resemblance of SN~2009hd to
other SNe~II-L.

\subsection{Extinction}\label{specph_extinc}

To derive consistent estimates of the extinction, $A_V$, toward SN
2009hd, we adopted two different methods, all based on the comparison
of the SN's spectral energy distribution (SED) and luminosity
with those of other ``standard'' CC-SNe. The estimates we derive below
relate to the SN at early times. There is some hint, from both the
color curves and spectrum, that the extinction decreased at late
times; however, as discussed above, what we consider the more likely explanation is contamination of the SN light by neighboring stars.

(i) We matched simultaneously the absolute $(B-V)_0$, $(V-R)_0$, $(R-I)_0$, and
$(V-I)_0$ colors curves of SN~2009hd with those of other SNe~II-L (see
Figure \ref{fig_color}). The best match is for $A_V = 3.9 \pm 0.1$
mag. This comparison was primarily weighted by the larger number of
data points at early epochs.

(ii) The observed extinction was estimated 
by comparing the early-time optical SED of SN~2009hd with those of SNe~1990K, 2001cy,
and 2009kr at similar epochs. Spectra of the comparison SNe were first
corrected for redshift and extinction, and scaled to the distance of
SN~2009hd. The average of the good matches in all cases is $A_V = 3.7
\pm 0.2$ mag.

We find that these
extinction estimates from the two quite similar
methods agree
reasonably well with each other. Hence, we adopt $A_V^{\rm tot} =
3.80 \pm 0.14$ mag as the extinction toward
SN~2009hd. The foreground Galactic component to this extinction is a relatively low 
0.11 mag \citep{schlegel98}, so, therefore, most of the extinction must be local to the SN.

This extinction estimate is consistent with the constraint on the reddening
we can derive from the Magellan+MagE spectrum. We measured the equivalent width (EW) of the
Na\,{\sc i}~D line at the host-galaxy redshift ($z_0 = 0.0024$) and
found for the two line components,
EW(Na\,{\sc i}~D1 $\lambda$5890) = $1.25 \pm 0.05$~\AA\ and
EW(Na\,{\sc i}~D2 $\lambda$5896) = $1.10 \pm 0.05$~\AA. Considering
that EW(Na\,{\sc i}~D1)$_{\rm hr}$ + EW(Na\,{\sc i}~D2)$_{\rm hr}$ =
EW(Na\,{\sc i}~D)$_{\rm lr}$ (Elias-Rosa et al., in
prep.), where ``hr'' and ``lr'' denote ``high'' and ``low
resolution,'' respectively, this results in 
EW(Na\,{\sc i}~D) = $2.35 \pm 0.07$. 
However, since the ratio between the Na\,{\sc i}~D component
lines is close to saturation (1.14, with saturation at 1.1;
\citealt{spitzer48}), and given the considerable scatter in the distribution of SNe in the EW(Na\,{\sc i}~D) vs.~$E(B-V)$ plane (e.g., see \citealt{eliasrosa07}\footnote[15]{http://pos.sissa.it/cgi-bin/reader/conf.cgi?confid=60.} 
or \citealt{poznanski11}) and the large variance in this relation found in the literature,
the estimated value of ${E(B-V)}_{\rm tot}$ can
only be considered as a {\it lower limit}. For example, using the relation between
reddening and EW(Na\,{\sc i}~D) (for all type of SNe) from 
\citet[][$E(B-V) = 0.11\, \times$ EW(Na\,{\sc i}~D) $- 0.01$]{eliasrosa07}, 
and assuming the Cardelli et al. (1989) reddening law, with updated bandpass
wavelengths, and a Galactic foreground $E(B-V) = 0.03$ mag
\citep{schlegel98}, we estimate
${E(B-V)}_{\rm tot} \gtrsim 0.29 \pm
0.03$ mag ($A_V \gtrsim 0.90 \pm 0.09$ mag and ${E[V-I]}_{\rm tot}
\gtrsim 0.35 \pm 0.06$ mag). If instead, we use the relation given by \cite{olivares10} for 
SNe II-P ($E(B-V) \approx 0.25\, \times$ EW(Na\,{\sc i}~D)), we estimate
${E(B-V)}_{\rm tot} \gtrsim 0.62$ mag. In both cases, these lower limits are consistent with the
two direct extinction estimates above.

We note here the possible temporal evolution of the Na\,{\sc i}
feature. While it is strong at early times, indicative of the high
extinction toward the SN, the feature's strength appears to
be reduced by
$\sim 20\%$ (EW[Na\,{\sc i}~D] = $1.87 \pm 0.05$) at late times. Such
variability in the Na\,{\sc i}~D lines has been observed before ---
for example, in the faint transient SN~2008S in NGC 6946
\citep{smith09,botticella09}, or in thermonuclear SNe, such as SN
2006dd \citep{stritzinger10}, and SN 1999cl and SN 2006X (see
\citealt{blondin09}, and references therein). However, in these latter two cases,
the EW of the lines increased with time, rather than
decreased. Considering the expected recombination of Na with time, such a variation for SN 2009hd, 
if real, may be due to ionization changes
in the CSM, as a result of the weaker radiation field from the SN at
later times. This effect could also be related to the possible decline
in the local extinction mentioned in \S \ref{specph_spec}. On the
other hand, another, possibly more likely, explanation for the apparent
change in the Na\,{\sc i}~D EW is dilution of this spectral feature by light from nearby stars and H\,{\sc ii} regions, as previously noted.


\section{Identification of the Progenitor Candidate}\label{identification}

Shortly after the discovery of SN~2009hd, we isolated 12
archival\footnote[16]{http://archive.stsci.edu/hst/.} {\sl
  HST\/}/WFPC2 CR-SPLIT image pairs of M66 in F555W ($\sim V$) and 5
pairs in F814W ($\sim I$), in order to identify any stars in the images
that could be candidates for the SN progenitor. These images were obtained between 1997
November and 1998 January by program GO-6549 (PI: A.~Sandage) and used
originally to derive a Cepheid-based distance to M66
\citep{saha99}. The exposure times in each band were
2400--2500~s. Combining the images in each band over all epochs of the
Cepheid observations yields a total image depth of $\sim 16$~hr in
F555W and $\sim 7$~hr in F814W. The combination of the pre-explosion
images for each band scaled to the PC chip pixel scale ($0{\farcs}045$
pix$^{-1}$), using the {\it drizzle\/} algorithm \citep{fruchter02},
is available from the {\sl HST\/} Legacy
Archive\footnote[17]{http://archive.stsci.edu/hlsp/.}. The input images in the 
{\it drizzle\/} algorithm are weighted according to the statistical significance of each pixel, and 
{\it drizzle\/} removes the effects of 
geometric distortion on both image shape and photometry. In addition, it combines dithered 
images in the presence of cosmic rays and improves the resolution of the mosaic. Thus, the use of 
the drizzled image mosaics for the identification of a progenitor candidate is ideal.

Photometry of the post-explosion {\sl HST\/} images was performed
using the DOLPHOT\footnote[18]{The ACS module of DOLPHOT is an
  adaptation of the photometry package HSTphot. We used v1.1, updated 2010 January 6, from
  http://purcell.as.arizona.edu/dolphot/.} package especially designed
for ACS. The post-explosion images were all obtained at the same epoch
in each band and with a small dithered offset between exposures. Thus,
we measured these relative offsets with respect to one fiducial image
before running DOLPHOT. The output from the package automatically
includes the transformation from flight-system F555W and F814W to the
corresponding Johnson-Cousins \citep{bessell90} magnitudes (in $V$ and
$I$), following the prescriptions of \citet{sirianni05}. We include
photometry of the SN in Table~\ref{table_ph}.

\begin{figure*}
\epsscale{1.0} 
\plotone{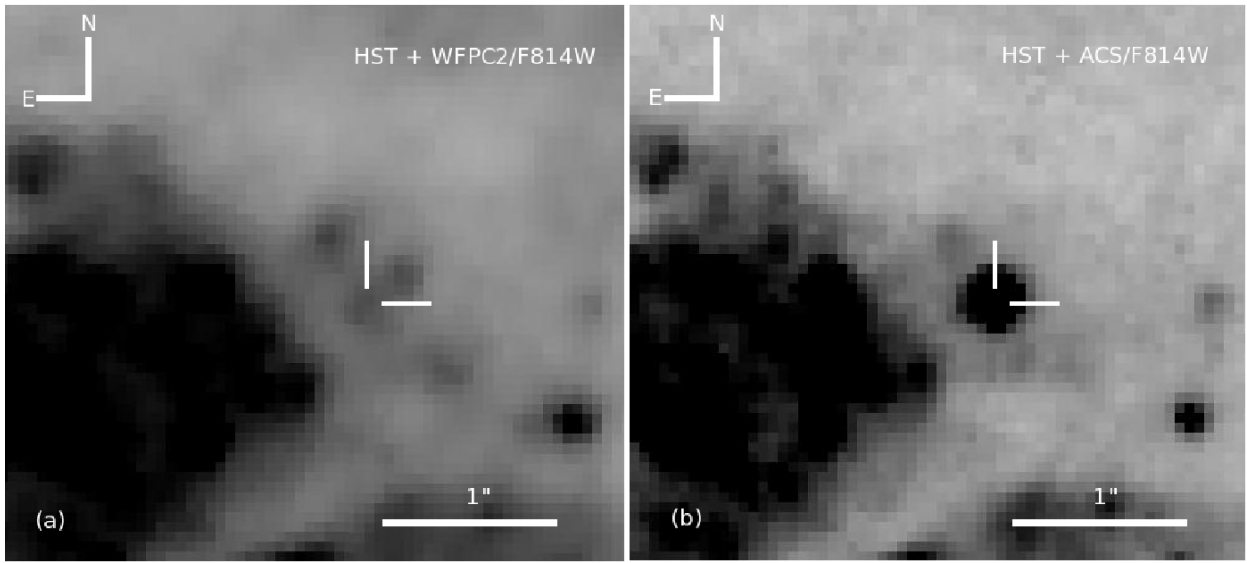}
\caption{(a) Subsections of the pre-explosion {\sl HST\/} WFPC2/F814W
  images of M66, and (b) the post-explosion {\sl HST\/} ACS/F814W
  image of SN~2009hd. The positions of the candidate
  progenitor and the SN are indicated by tick marks. }
\label{fig_progenitor}
\end{figure*}

For the SN progenitor candidate identification, 
it was necessary that we use high-resolution images of the SN, ideally at the
same or better resolution as the pre-SN {\sl HST\/}
image. Accordingly, we obtained {\sl HST\/} Advanced Camera for
Surveys (ACS; Wide Field Channel, $\sim 0{\farcs}05$ pix$^{-1}$)
images in F555W and F814W on 2009 December 14, as part of our
Target-of-Opportunity program GO-11575 (PI: S.~Van Dyk). The SN was still quite bright
at the time of these observations (see Table~\ref{table_ph}), although
it was not saturated in the images. The observations in each band were
in pairs of dithered 260~s exposures.  These individual exposures were
{\it drizzled} to produce final mosaics in each band. 
A comparison of a {\sl HST\/} SN image to a pre-SN image is shown in 
Figure \ref{fig_progenitor}.

We achieved high-precision relative astrometry between the pre-explosion 
drizzled WF2 image in each band (the SN site is located on 
this chip of the pre-explosion images; this image is of the same depth as the drizzled
full mosaic image) and the post-explosion ones by geometrically
transforming the former to match precisely the latter.
Using the IRAF task {\it geomap}, we carried out several geometric
transformations between the two sets of images, using different numbers of point-like
sources in common between the two datasets (from $\sim$10--30 manually selected sources
to $\gtrsim$ 300 sources using an automated matching algorithm), all 
with root-mean-square (rms) 
uncertainty $\lesssim 0{\farcs}15$. 
We then transformed the pre-explosion images to the post-explosion ones 
using the IRAF task {\it geotran\/}. 
The positions (and their
uncertainties) of the SN and the progenitor candidate were generally obtained
by averaging the measurements from two centroiding methods, the task 
{\it daofind\/} within IRAF/DAOPHOT and {\it imexamine} within IRAF. 
For all solutions, the resulting SN position in the pre-explosion images in the two bands 
were all consistent.

\begin{figure*}
\epsscale{1.0} 
\plotone{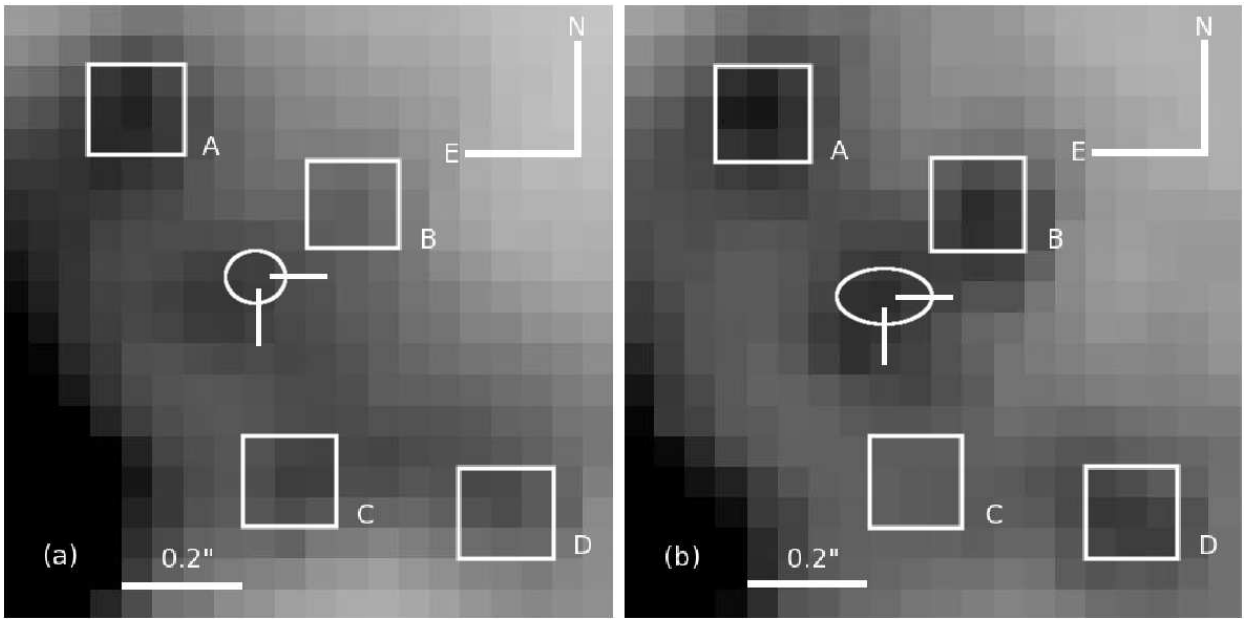}
\caption{Subsections of the pre-explosion {\sl HST\/} WFPC2 images of M66 in
  (a) F555W and (b) F814W. The position of the SN is indicated by a 5$\sigma$
  positional uncertainty ellipse ($0{\farcs}04 \times 0{\farcs}03$ for
  the F555W image and $0{\farcs}08 \times 0{\farcs}04$ for the F814W image). The four objects ``A,'' ``B,'' ``C,''
  and ``D'' were not used in the image astrometry, due to their
  relative faintness; they are labeled here only to aid the eye in locating
  the progenitor in the field.}
\label{fig_progenitor_comp}
\end{figure*}

As a result, in the pre-SN F814W image we identify an object very 
near the SN position which we consider to be the progenitor candidate.
It is not detected at this position in the F555W image; 
see Figure~\ref{fig_progenitor_comp}.
The average final pixel
position in the pre-SN F814W image is [1074.82, 1634.13] for the SN, and [1074.43, 1634.10] for 
the candidate in the transformed SN image. The candidate is one of the detected
stars near the edge of the pre-explosion error circle suggested by \citet{maoz09} in his initial 
attempt to identify the SN progenitor. The differences between the SN and the
progenitor candidate positions,
compared with the total estimated
uncertainty in the astrometry, are given in Table
\ref{table_error}. This latter uncertainty was calculated as a
quadrature sum of the uncertainties in the SN and progenitor candidate
positions, and the rms uncertainty in the geometric transformation.

From the results in Table \ref{table_error}, it can be seen that the
difference between the SN position and the position of
the progenitor candidate in F814W is larger in right ascension than the 
measurement uncertainties. This offset could be due to the proximity of
one or more sources to the progenitor candidate position, which may have affected 
the measurement of the object's centroid, or to the difficulty in making accurate centroid
measurements, given the complex background in the bright spiral arm of
the host galaxy on which the progenitor candidate is located. In Figure~\ref{fig_progenitor_comp}
one can see that the reference object, labelled ``B" ($\sim 0{\farcs}25$ northwest of the candidate),
is visible in the F814W image, but is essentially invisible in the F555W image. The opposite is
true for the object, labelled ``C" ($\sim 0{\farcs}33$ southwest of the candidate). We 
emphasize that no other point-like source can be isolated within a 1.5$\sigma$ (i.e., $\leq$2 
pixel) radius from the progenitor candidate position (see Figures \ref{fig_progenitor} and 
\ref{fig_progenitor_comp}). The alternative possibility is that we have not detected the 
progenitor for SN~2009hd. We consider both of these possibilities below. 


\section{The Nature of the Progenitor}\label{discussion}

Next, we attempted to measure the brightness of the progenitor candidate. 
We performed photometry on each of the individual pre-explosion {\sl HST\/} 
images in both bands using
HSTphot\footnote[19]{HSTphot is a stellar photometry package designed for use with WFPC2 
images \citep{dolphin00}. We used v1.1.7b, updated 2008 July 19.}. (HSTphot is not
designed to run on the drizzled image mosaics.) 
We ran the package on the CR-SPLIT pairs for both bands, 
with option flag 10, which combines turning on
the local sky determination, turning off empirically determined
aperture corrections (using default values instead), and turning on
PSF residual determination, with a total detection threshold of
$3\sigma$. As a result, we obtained $m_{\rm F814W} = 23.54 \pm 0.14$ mag for the progenitor 
candidate.
HSTphot did not detect in F555W the candidate 
detected in F814W, even with the detection threshold set as low as $1\sigma$. To be sure that
this object is truly not detected in F555W, we repeated the measurements, varying the
number and order of the images for each filter input into HSTphot (we note that the images in 
each band can be split into two subsamples, depending on the various camera pointings), 
both automatically and forcing HSTphot to make a detection at the centroid position of the object
in F814W. Complicating the detection is the background. The number of counts around the SN
position in the F555W pre-SN images is nearly a factor of two higher than in a similarly sized 
region measured in a ``empty" area, relatively free of galaxy contamination.

We note that HSTphot reported the progenitor
candidate as having an ``object type'' flag of ``1,'' indicating that the
source is likely stellar. Additionally, the object had a ``sharpness"
range of about $-0.20$ to 0.16 (which indicates the reliability that a
detected source is indeed point-like; \citealt{dolphin00}). A ``good
star'' value for this parameter is between $-0.3$ and +0.3
\citep{leonard08}. Finally, the ``$\chi$" value is $\sim 1.5$; hence,
it is reasonable to call the candidate a ``good and clean'' star
($\chi < 1.5$, see \citealt{dolphin00}). We therefore consider it very
likely that the candidate is a single star.

\begin{figure*}
\epsscale{1.0}
\centering
\includegraphics[height=5truein,width=3.8truein,angle=-90]{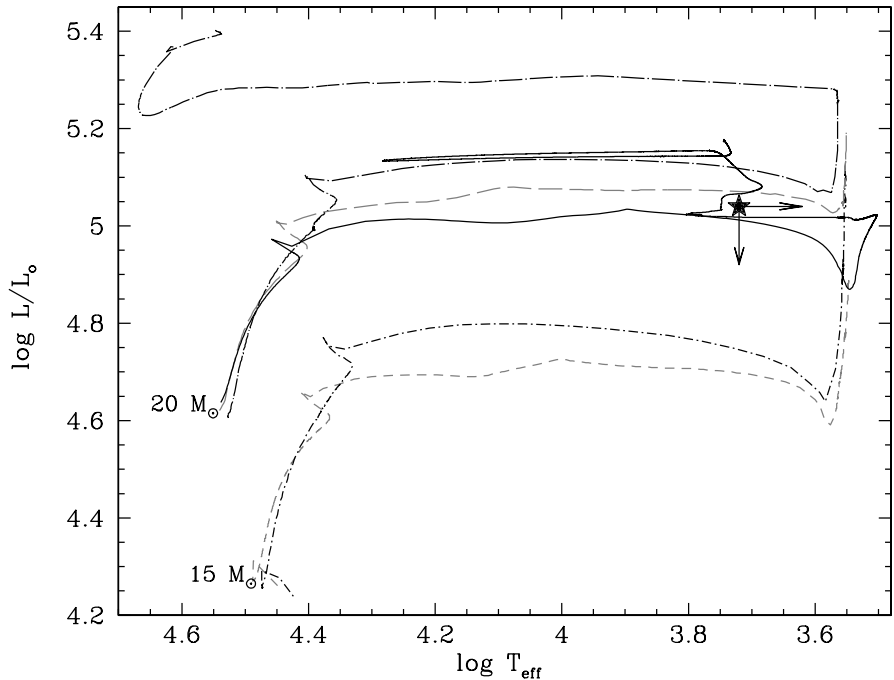}
\caption{A Hertzsprung-Russell diagram ($T_{\rm eff}$ (K) vs. $L_{\rm bol}$ (L$_{\sun}$)) showing the upper limit considering a contaminated progenitor ({\it filled 5-pointed star}). Model stellar evolutionary tracks for solar metallicity are also illustrated for $M_{\rm ini}$ of 15 and 20 M$_{\sun}$ without rotation ({\it short-dashed}, and {\it long-dashed gray lines}) and with rotation of $v_{\rm ini} = 300$ \kms\ ({\it dot-short-dashed}, and {\it dot-long-dashed black lines}; \citealt{hirschi04}); and for 20 M$_{\sun}$ considering pulsation-induced mass loss enhancement ({\it solid black line}; \citealt{yoon10}).
\label{fig_hrd}}
\end{figure*}

Of course, given the offset of the SN position from this star, we must still take into account
the possibility that we have not actually detected the progenitor in either band. 
We therefore attempted to constrain the brightness of an undetected progenitor by adding an 
artificial star at the precise SN position (see \S~\ref{identification}) in the pre-SN images for both bands. 
The PSF for the input artificial star was created from bright, isolated stars seen in the drizzled 
image mosaic in each band using IRAF/DAOPHOT. The brightness of the artificial star was
varied, until it was practically not detectable above the background. Then, the instrumental 
magnitudes of the artificial star were measured in the images using PSF fitting, and 
subsequently calibrated to the flight-system photometric
system using the magnitude and colors of 8 brighter stars in the vicinity of the SN, measure with HSTphot in the {\sl HST\/} WFPC2 images. We find
that $>$ 26.1 mag in F555W and $>$ 23.5 mag in F814W. Note that the limit in F814W is very similar to the measure brightness of the progenitor candidate in this band.
Correcting by the same total extinction as assumed for the SN ($A_V^{\rm tot} = 3.80 \pm 0.14$ mag; see \S~\ref{specph_extinc}) and the adopted distance modulus \citep[$\mu = 29.86 \pm 0.08$ mag;][]{freedman01}, we find
a limit on the absolute magnitude of an undetected progenitor, $M_V^0 \gtrsim -7.6$ mag. 
Furthermore, assuming a bolometric correction of
$-0.29$ mag for supergiants at solar metallicity (\citealt{kurucz93}; Kurucz Atlas 9 
models, CD-ROMs 13, 18; see below for the metallicity measurement in the SN
environment) appropriate for $(V-I)_0 \gtrsim 0.99$ mag (this is the limit on the color that arises 
from the reddening-corrected upper limits at $V$ and $I$, above), we find an
upper limit to such a star's bolometric luminosity of
$L_{\rm bol} \lesssim 10^{5.04}\ {\rm L}_{\sun}$. The limit on the effective
temperature, from the color limit, is then $T_{\rm eff} \lesssim 5200$ K.
We place this limit on the progenitor in the Hertzsprung-Russell (HR) diagram
in Figure~\ref{fig_hrd}. 

We can compare the limit shown in the figure to massive-star evolutionary models 
without rotation and with rotation \citep[$v_{\rm ini} = 300$ \kms][]{hirschi04}. One can see
that the appropriate initial masses to consider, of those that are available for these models, are 15 and 20 M$_{\sun}$. (We note that the majority of stars do not rotate as fast as those in the
models used here; e.g., less than about 10$\%$ of young Galactic
O-type stars have rotation speeds of 300 km s$^{-1}$ or more;
\citealt{penny04}.) We also show the evolutionary tracks for a model star of initial mass 20 M$_{\sun}$ with 
pulsationally enhanced mass loss \citep{yoon10}. We assume that these tracks correspond
to solar metallicity. We determined directly the metallicity in the SN environment from a
neighboring H\,{\sc ii} region detected in the late-time SN spectrum (see \S~\ref{specph_spec}).
Measuring the H$\alpha$ and [N~II] $\lambda$6584 line
intensities and applying a cubic-fit relation between the ratio of
these two lines and the oxygen abundance \citep{pettini04}, we find 12
+ log(O/H) = $8.43 \pm 0.05$, which is somewhat lower than the solar
abundance, $8.69 \pm 0.05$ \citep{asplund09}.  We adopt the
criterion given by \cite{smartt09}, which, taking into account the uncertainties in
the different metallicity estimates, considers 12 + log(O/H) = 8.4 as 
the dividing line between solar and subsolar metallicity evolutionary tracks. 

To convert this progenitor detection limit to an upper limit on the
initial mass of the progenitor, we considered the luminosities of the
evolutionary tracks in the post-red supergiant (RSG) phase (see Figure \ref{fig_hrd}).
We made this estimation three different ways. First, we compared the 
luminosity upper limit to the stellar {\em endpoints\/} of the
evolutionary tracks, as \citet{fraser09} did in the case of the
SN~II-L~2009kr. Doing so, we find that $M_{\rm ini} \lesssim\ 20 M_{\sun}$
for the models with no rotation, and $\lesssim 15\ M_{\sun}$ for the
models with rotation. If we, instead, consider where the upper limit intersects the
evolutionary tracks, we find a similar upper initial mass limit,
$\lesssim\ 20 M_{\sun}$, for the non-rotating models, and $\lesssim 18\
M_{\sun}$ for the models with rotation. If we consider the model with
pulsationally-induced mass loss at $20\ M_{\sun}$, one can see that it
naturally achieves terminus blueward of the RSG phase. \citet{yoon10} found
that their $17\ M_{\sun}$ model only experiences pulsations after Ne
burning. One can imagine, then, that, as the models approach $20\
M_{\sun}$ from that lower mass, that pulsations become a stronger effect,
and the models also terminate toward the blue, rather than the red. Thus,
conservatively, we can say that, if pulsations are important for these
massive stars, the luminosity limit could correspond to a mass limit 
$\lesssim\ 20\ M_{\sun}$; however, more liberally, this can extend to
somewhat a lower mass limit of $\lesssim\ 18$--$19\ M_{\sun}$.

\section{Discussion and Conclusions}\label{conclusions}

Observations at early times and at later (nebular) phases suggest that
SN 2009hd is a SN~II-L, similar to SN~1990K \citep{cappellaro95} and
SN~2009kr \citep{eliasrosa10}. However, the lack of data between $\sim
30$ and 100~d leaves open the possibility that SN~2009hd may be a hybrid between SNe~II-L and SNe~II-P. If it is a SN~II-L, it is one of
the most extinguished examples detected thus far,  
possibly as a result of proximity to a dust lane that is clearly evident in pre-SN {\sl HST\/}
images. Projection effects may further play a role, given the
inclination of M66 \citep[65$\arcdeg$;][]{tully88}.

Comparing {\sl HST\/} images of the host galaxy prior to explosion and
of the SN at late times, the resulting high-precision relative astrometry has
provided us with a means of identifying the SN progenitor. 
A progenitor candidate is possibly identified in the available F814W, but 
definitely not in the F555W, pre-SN images.
Even with the high spatial resolution of the {\sl HST\/} WFPC2 images, the uncertainties in 
the astrometry are large enough that we are unable to make a definitive identification.
Given the relative proximity of the host galaxy to us (9.4 Mpc), this is rather surprising. The high extinction, combined with the complex background in the SN environment, likely
lead to this level of uncertainty.

We can constrain the properties (luminosity, effective temperature) of the progenitor star for
SN~2009hd. Although the inferred properties are certainly consistent with a red supergiant (RSG),
the constraints also allow for the possibility that the star is more yellow than red. A yellow color could imply that it exploded in a post-RSG evolutionary state (or perhaps that it is an RSG in a binary system with a bluish companion). The progenitor may have evolved back to the blue after reaching
the RSG stage, for example, because it has lost mass (e.g., \citealt{podsiadlowski92,meynet11}) as a result of rotationally induced winds (e.g., \citealt{hirschi04}), or through pulsationally enhanced mass loss (e.g., \citealt{yoon10}). The fact that the H lines are weak in the nebular spectrum may indicate that the star exploded with a
low-mass H envelope, presumably lost through high mass loss via stellar winds prior to the termination of He burning. Consequently, the star would evolve toward the blue before its terminus (e.g., \citealt{meynet11}, and references therein). Through comparison with high-mass stellar evolutionary tracks at solar metallicity, we are able, conservatively, to constrain the initial mass of the progenitor to $M_{\rm ini} \lesssim 20\ {\rm M}_{\sun}$.

This limit to the initial mass estimate for the SN~2009hd progenitor is consistent with the range
in mass found for the SN~II-L~2009kr, $M_{\rm ini} \approx 18$--24 M$_{\sun}$ 
\citep[][see also \citealt{fraser09}]{eliasrosa10}. The progenitor of SN~2009kr was inferred likely to be a yellow supergiant. The progenitor of the unusual, high-luminosity SN~II-P~2008cn also appears to have been a
yellow supergiant, with $M_{\rm ini} =15 \pm 2\ {\rm M}_{\sun}$ \citep{eliasrosa09}.
SNe 2009hd, 2009kr, and 2008cn may be a bridge between the RSGs that explode as
SNe~II-P and the high-mass, luminous blue variables or more extreme
RSGs that may explode as SNe~IIn \citep{smith10}. 

Ultimately, a very-late-time set of multi-band
images of the SN~2009hd field should be obtained with {\sl HST\/} and the Wide Field Camera 3 (WFC3) many years later, when the SN has greatly faded, as has been done for the SN~IIn~2005gl \citep{galyam09} and for the SN~IIb 1993J and SN~II-P 2003gd \citep{maund09},
to better distinguish the constituents of the SN~2009hd environment and help decipher
the true nature of its progenitor.


\acknowledgments

This research is based in part on observations made with the NASA/ESA
{\it Hubble Space Telescope}, obtained from the Data Archive at the
Space Telescope Science Institute, which is operated by the
Association of Universities for Research in Astronomy (AURA), Inc.,
under NASA contract NAS 05-26555; the SMARTS Consortium 1.3~m
telescope located at Cerro Tololo Inter-American Observatory (CTIO),
Chile; the 1.5~m telescope located at Palomar Observatory; the
Liverpool Telescope operated on the island of La Palma by Liverpool
John Moores University in the Spanish Observatorio del Roque de los
Muchachos of the Instituto de Astrofisica de Canarias with financial
support from the UK Science and Technology Facilities Council; the
Nordic Optical Telescope, operated on the island of La Palma jointly
by Denmark, Finland, Iceland, Norway, and Sweden, in the Spanish
Observatorio del Roque de los Muchachos of the Instituto de
Astrofisica de Canarias; the 6.5~m Magellan Clay and Baade Telescopes
located at Las Campanas Observatory, Chile; and the W. M. Keck
Observatory, which is operated as a scientific partnership among the
California Institute of Technology, the University of California, and
NASA, with generous financial support from the W. M. Keck Foundation.
We are grateful to the staffs at these observatories for their excellent
assistance with the observations.

N.E.R. thanks Stefano Benetti and Alberto Noriega-Crespo for useful
discussions and Avet Harutyunyan for his help. Many of the spectra
used here for comparison were obtained from the Padova-Asiago
Supernova Archive (ASA). This work has made use of the NASA/IPAC
Extragalactic Database (NED), which is operated by the Jet Propulsion
Laboratory, California Institute of Technology, under contract with
NASA.  This work has been supported by the MICINN grant
AYA08-1839/ESP, by the ESF EUROCORES Program EuroGENESIS (MICINN grant
EUI2009-04170), by SGR grants of the Generalitat de Catalunya, and by
EU-FEDER funds.

The research of A.V.F.'s supernova group has been generously supported
by National Science Foundation grant AST-0908886 and the TABASGO
Foundation, as well as by NASA through grants AR-11248, AR-12126, and
GO-11575 from the Space Telescope Science Institute, which is operated
by AURA, Inc., under NASA contract NAS 5-26555.

{\it Facilities:} \facility{HST (WFPC2)}, \facility{HST (ACS)},
\facility{CTIO:1.3m(SMARTS)}, \facility{PO:1.5m}, \facility{PO: 5m
  (TripleSpec)}, \facility{Liverpool:2m (RATCam)}, \facility{NOT
  (ALFOSC)}, \facility{Magellan: Baade (IMACS)}, \facility{Magellan:
  Clay (MagE)}, \facility{Keck-II: DEIMOS}.


\clearpage

\clearpage

\begin{deluxetable}{lcccc}
\tablewidth{0pt}
\tablecaption{Adopted Magnitudes for the Local Sequence Stars in Figure \ref{fig_seq} \label{table_seq} }
\tablehead{ \colhead{Star} &
\colhead{$B$}    & \colhead{$V$}  &
\colhead{$R$}  & \colhead{$I$}  \\
\colhead{} & \colhead{(mag)}    & \colhead{(mag)}  &
\colhead{(mag)}  & \colhead{(mag)} } 
\startdata
1 & 13.72$\pm$0.01 & 13.17$\pm$0.01 & 12.83$\pm$0.01 & 12.50$\pm$0.01 \\
2 & 16.26$\pm$0.01 & 15.47$\pm$0.01 & 14.97$\pm$0.01 & 14.49$\pm$0.03 \\
3 & 17.16$\pm$0.01 & 16.54$\pm$0.01 & 16.16$\pm$0.02 & 15.77$\pm$0.01 \\
4 & 16.77$\pm$0.01 & 16.24$\pm$0.01 & 15.92$\pm$0.01 & 15.55$\pm$0.01 \\
5 & 16.56$\pm$0.01 & 15.95$\pm$0.01 & 15.59$\pm$0.01 & 15.18$\pm$0.01 \\
6 & 18.03$\pm$0.01 & 16.49$\pm$0.01 & 15.41$\pm$0.01 & 13.49$\pm$0.24 \\
\enddata
\end{deluxetable}

\begin{deluxetable}{lcccccl}
\tablewidth{0pt}
\tablecaption{Photometry of SN~2009hd \label{table_ph}} 
\tablehead{ \colhead{Date} &
\colhead{JD} & \colhead{$B$}    & \colhead{$V$}  &
\colhead{$R$}  & \colhead{$I$} & \colhead{Source} \\
\colhead{} & \colhead{$-$2,400,000.00} & \colhead{(mag)}    & \colhead{(mag)}  &
\colhead{(mag)}  & \colhead{(mag)} & \colhead{}} 
\startdata
06/07/09 & 55019.48 &  18.25$\pm$0.05 &  16.73$\pm$0.06 &  15.82$\pm$0.05 &  14.66$\pm$0.04 &  CTIO \\
08/07/09 & 55021.46 &  18.43$\pm$0.04 &  16.78$\pm$0.03 &  15.83$\pm$0.04 &  14.69$\pm$0.04 &  CTIO \\
09/07/09 & 55021.70 &  18.44$\pm$0.06 &  16.75$\pm$0.03 &  15.75$\pm$0.04 &  \nodata &  P60 \\
10/07/09 & 55022.70 &  18.56$\pm$0.05 &  16.78$\pm$0.04 &  15.81$\pm$0.02 &  14.68$\pm$0.04 &  P60 \\
10/07/09 & 55023.45 &  18.41$\pm$0.07 &  16.75$\pm$0.02 &  15.76$\pm$0.03 &  14.71$\pm$0.03 &  CTIO \\
12/07/09 & 55025.46 &  18.57$\pm$0.08 &  16.83$\pm$0.06 &  15.79$\pm$0.06 &  14.69$\pm$0.04 &  CTIO \\
14/07/09 & 55026.69 &  18.65$\pm$0.03 &  \nodata &  15.82$\pm$0.06 &  \nodata &  P60 \\
16/07/09 & 55029.46 &  18.73$\pm$0.07 &  16.84$\pm$0.02&  15.83$\pm$0.05 &  14.83$\pm$0.03 &  CTIO \\
18/07/09 & 55031.46 &  18.98$\pm$0.06 &  16.94$\pm$0.02&  15.94$\pm$0.05 &  14.81$\pm$0.03 &  CTIO \\
24/11/09 & 55160.71 &  \nodata &  20.64$\pm$0.05 &  19.63$\pm$0.02 &  18.75$\pm$0.02 &  LT \\
14/12/09 & 55179.58 &  \nodata &  20.87$\pm$0.01 &  \nodata &  18.95$\pm$0.01 &  HST \\
19/12/09 & 55184.86 &  \nodata &  21.09$\pm$0.10 &  20.11$\pm$0.08 &  \nodata &  Mag \\ 
20/01/10 & 55216.55 &  \nodata &  20.85$\pm$0.04 &  20.30$\pm$0.05 &  19.54$\pm$0.04 &  LT \\
15/02/10 & 55242.93 &  \nodata &  \nodata &  20.83$\pm$0.05 &  19.97$\pm$0.07 &  Keck \\
19/03/10 & 55274.54 &  \nodata &  21.43$\pm$0.09 &  20.99$\pm$0.08 &  20.33$\pm$0.13 & NOT \\
\enddata
\tablecomments{CTIO = CTIO 1.3~m SMARTS telescope + ANDICAM; P60 =
  Palomar 1.5~m telescope + CCD camera; LT = 2.0~m Liverpool telescope
  + RATCam; HST = Hubble Space Telescope + ACS camera; Mag = 6.5~m
  Baade Magellan Telescope + IMACS; Keck = 10~m Keck-II telescope +
  DEIMOS; NOT = 2.6~m Nordic Optical Telescope + ALFOSC.}
\end{deluxetable}

\begin{deluxetable}{lcc}
\tablewidth{0pt}
\tablecaption{SN and Progenitor Candidate Position Comparison\label{table_error}}
\tablehead{
\colhead{Angular Quantity}           & \colhead{$V$ ($\alpha/\delta$)}      &
\colhead{$I$ ($\alpha/\delta$)}
}
\startdata
Uncertainty in the progenitor position (mas) & \nodata & 14/5\\
Uncertainty in the SN position (mas) & 1/3  & 1/3\\
Geometric transformation (mas) & 8/5 & 7/6\\
Total uncertainty (mas) & 8/6 & 15/8\\
\tableline
Difference in position (mas) & \nodata & 20/1\\
\enddata
\tablecomments{Uncertainties in the SN and candidate position, in
  milliarcsec (mas), were estimated as the standard deviation of the
  average. Geometric transformation errors are derived from the
  positional differences between the fiducial stars used in the
  transformation. The total uncertainty is the quadrature sum of these
  uncertainties. The last line lists the residual difference between
  the SN and progenitor position after the geometric transformation.}
\end{deluxetable}


\begin{thebibliography}{}

\bibitem[Asplund et al.(2009)]{asplund09}
  Asplund, M., Grevesse, N., Sauval, A.~J., \& Scott, P. 2009, \aapr,
  47, 481
\bibitem[Barbon, Ciatti, \& Rosino(1979)]{barbon79} Barbon, R.,
  Ciatti, F., \& Rosino, L. 1979, \aap, 72, 287
\bibitem[Barbon, Ciatti, \& Rosino(1982)]{barbon82} Barbon, R.,
  Ciatti, F., \& Rosino, L.\ 1982, \aap, 116, 35
\bibitem[Barbon et al.(1995)]{barbon95} Barbon, R., et al. 1995,
  \aap, 110, 513
\bibitem[Baron et al.(2000)]{baron00} Baron, E., et al. 2000, \apj,
  545, 444
\bibitem[Benetti et al.(1998)]{benetti98} Benetti, S., et
  al. 1998, \mnras, 294, 448
\bibitem[Berger, Foley, \& Covarrubias(2009)]{berger09} Berger, E.,
  Foley, R.~J., \& Covarrubias, R. 2009, The Astronomer's Telegram
  2118, 1
\bibitem[Bessell(1990)]{bessell90} Bessell, M.~S. 1990, \pasp, 102,
  1181
\bibitem[Blondin et al.(2009)]{blondin09} Blondin, S., et al. 2009,
  \apj, 693, 207
\bibitem[Botticella et al.(2009)]{botticella09} Botticella, M.~T., et
  al. 2009, \mnras, 398, 1041
\bibitem[Buta(1982)]{buta82} Buta, R.~J.\ 1982, \pasp, 94, 578
\bibitem[Cappellaro et al.(1990)]{cappellaro90} Cappellaro, E., Della
  Valle, M., Iijima, T., \& Turatto, M.\ 1990, \aap, 228, 61
\bibitem[Cappellaro et al.(1995)]{cappellaro95} Cappellaro, E.,
  Danziger, I.~J., Della Valle, M., Gouiffes, C., \& Turatto, M.\ 1995,
  \aap, 293, 723
\bibitem[Cardelli, Clayton, \& Mathis(1989)]{cardelli89} Cardelli,
  J. A., Clayton, G. C., \& Mathis, J. S. 1989, \apj, 345, 245
\bibitem[Ciatti \& Rosino(1977)]{ciatti77} Ciatti, F., \& Rosino,
  L. 1977, \aap, 56, 59
\bibitem[Clocchiatti et al.(1996)]{clocchiatti96} Clocchiatti, A., et
  al. 1996, \aj, 111, 1286
\bibitem[Dessart \& Hillier(2005)]{dessart05} Dessart, L., \& Hillier,
  D. J. 2005, \aap, 437, 667
\bibitem[Dolphin(2000)]{dolphin00} Dolphin, A.~E. 2000, \pasp, 112,
  1383
\bibitem[Elias et al.(2007)]{eliasrosa07} Elias, N., Beckman, J. E.,
  Benetti, S., Cappellaro, E., \& Turatto, M. 2007, in
Supernovae: Lights in the Darkness (http://pos.sissa.it/cgi-bin/reader/conf.cgi?confid=60)
\bibitem[Elias-Rosa et al.(2009)]{eliasrosa09} Elias-Rosa, N., et
  al.\ 2009, \apj, 706, 1174
\bibitem[Elias-Rosa et al.(2010)]{eliasrosa10} Elias-Rosa, N., et
  al.\ 2010, \apj, 714, L254
\bibitem[Elmhamdi et al.(2003)]{elmhamdi03} Elmhamdi, A., et
  al.\ 2003, \mnras, 338, 939
\bibitem[Faber et al.(2003)]{faber03} Faber, S.~M., et al. 2003,
  Society of Photo-Optical Instrumentation Engineers (SPIE) Conference
  Series, 4841, 1657
\bibitem[Filippenko(1982)]{fil82} Filippenko, A.~V. 1982, \pasp, 94,
715
\bibitem[Filippenko(1997)]{filippenko97} Filippenko, A.~V.\ 1997,
  \araa, 35, 309
\bibitem[Foley et al.(2003)]{foley03} Foley, R.~J., et al. 2003,
  \pasp, 115, 1220
\bibitem[Foley et al.(2009)]{foley09} Foley, R.~J., et al. 2009, \aj,
  138, 376
\bibitem[Fraser et al.(2009)]{fraser09} Fraser, M., et al.\ 2010, \apj, 714, L280
\bibitem[Freedman et al.(2001)]{freedman01} Freedman, W.~L., et
  al. 2001, \apj, 553, 47
\bibitem[Fruchter \& Hook(2002)]{fruchter02} Fruchter, A.~S., \& Hook,
  R.~N. 2002, \pasp, 114, 144
\bibitem[Gal-Yam \& Leonard(2009)]{galyam09} Gal-Yam, A., \& Leonard,
  D.~C. 2009, \nat, 458, 865
\bibitem[Hamuy et al.(2001)]{hamuy01} Hamuy, M., et al. 2001, \apj,
  558, 615
  \bibitem[Herter et al.(2008)]{herter08} Herter, T.~L., et al.\ 2008, Society of Photo-Optical Instrumentation Engineers (SPIE) Conference Series, 7014, 30
\bibitem[Hirschi, Meynet, \& Maeder (2004)]{hirschi04} Hirschi, R.,
  Meynet, G., \& Maeder, A. 2004, \aap, 425, 649
\bibitem[Horne(1986)]{horne86} Horne, K. 1986, \pasp, 98, 609
\bibitem[Kasliwal, Sahu, \& Anupama(2009)]{kasliwal-et-al-09} Kasliwal,
  M.~M., Sahu, D.~K., \& Anupama, G.~C. 2009, CBET, 1874, 1
\bibitem[Kasliwal(2009)]{kasliwal09} Kasliwal, M.~M. 2009, ATel 2113
\bibitem[Kelson(2003)]{kelson03} Kelson, D.~D. 2003, \pasp, 115, 688
\bibitem[Kennicutt et al.(2003)]{kennicutt03} Kennicutt, R.~C., Jr.,
  et al. 2003, \pasp, 115, 928
\bibitem[Kurucz(1993)]{kurucz93} Kurucz,~R. L. 1993, CDROM 13, 18
\bibitem[Leonard et al.(2002)]{leonard02} Leonard, D.~C., et
  al. 2002, \pasp, 114, 35
\bibitem[Leonard et al.(2008)]{leonard08} Leonard, D.~C., et
  al. 2008, \pasp, 120, 1259
\bibitem[Li et al.(2011)]{li10} Li, W., et al. 2011, \mnras, 412, 1441
\bibitem[Maoz(2009)]{maoz09} Maoz, D. 2009, The Astronomer's Telegram
  2114, 1
\bibitem[Marshall et al.(2008)]{marshall08} Marshall, J.~L., et
  al. 2008, Society of Photo-Optical Instrumentation Engineers (SPIE)
  Conference Series, 7014, 169
\bibitem[Matthews et al.(2002)]{matthews02} Matthews, K., et
  al. 2002, \aj, 123, 753
\bibitem[Maund \& Smartt(2009)]{maund09} Maund, J.~R., \& Smartt,
S.~J.\ 2009, Science, 324, 486
\bibitem[Meynet et al.(2011)]{meynet11} Meynet, G., et al. 2011, Bulletin de la Societe Royale des Sciences de Liege, 80, 266
\bibitem[Monard(2009)]{monard09} Monard, L.~A.~G. 2009, CBET 1867, 1
\bibitem[Olivares E. et al.(2010)]{olivares10} Olivares E., F., et al.\ 2010, \apj, 715, 833
\bibitem[Pastorello et al.(2006)]{pastorello06} Pastorello, A., et
  al. 2006, \mnras, 370, 1752
\bibitem[Patat et al.(1994)]{patat94} Patat, F., et al. 1994, \aaps,
  98, 443
\bibitem[Penny et al.(2004)]{penny04} Penny, L.~R., et al. 2004,
  \apj, 617, 1316
\bibitem[Pettini \& Pagel(2004)]{pettini04} Pettini, M., \& Pagel,
  B.~E.~J. 2004, \mnras, 348, 59
\bibitem[Podsiadlowski(1992)]{podsiadlowski92} Podsiadlowski, P. 1992, \pasp, 104, 717
\bibitem[Poznanski et al.(2011)]{poznanski11} Poznanski, D., et al 2011, \mnras, 415, L81
\bibitem[Pozzo et al.(2006)]{pozzo06} Pozzo, M., et al. 2006, \mnras,
  368, 1169
\bibitem[Prieto et al.(2008)]{prieto08} Prieto, J.~L., et al. 2008,
  \apj, 681, L9
\bibitem[Saha et al.(1999)]{saha99} Saha, A., et al. 1999, \apj, 522,
  802
\bibitem[Schlegel, Finkbeiner, \& Davis(1998)]{schlegel98} Schlegel,
  D.~J., Finkbeiner, D.~P., \& Davis, M. 1998, \apj, 500, 525
\bibitem[Schlegel(1996)]{schlegel96} Schlegel, E.~M.\ 1996, \aj, 111,
  1660
\bibitem[Sirianni et al.(2005)]{sirianni05} Sirianni, M., et
  al. 2005, \pasp, 117, 1049
\bibitem[Smartt et al.(2009)]{smartt09} Smartt, S.~J., et al. 2009,
  \mnras, 395, 1409
\bibitem[Smith et al.(2009)]{smith09} Smith, N., et al. 2009, \apj,
  697, L49
\bibitem[Smith et al.(2011)]{smith10} Smith, N., et al. 2011, \mnras, 412, 1522
\bibitem[Spitzer(1948)]{spitzer48} Spitzer, L.,~Jr. 1948, \apj, 108,
  276
\bibitem[Stetson(1987)]{stetson87} Stetson, P.~B. 1987, \pasp, 99,
  191
\bibitem[Stritzinger et al.(2011)]{stritzinger10} Stritzinger, M., et
  al. 2011, \aj, 140, 2036
\bibitem[Swartz, Wheeler, \& Harkness(1991)]{swartz91} Swartz, D.~A.,
  Wheeler, J.~C., \& Harkness, R.~P. 1991, \apj, 374, 266
\bibitem[Taubenberger et al.(2011)]{taubenberger11} Taubenberger, S., et al. 2011, \mnras, 413, 2140
\bibitem[Tsvetkov(1983)]{tsvetkov83} Tsvetkov, D.~Y. 1983, Peremennye
  Zvezdy, 22, 39
\bibitem[Tully(1988)]{tully88} Tully, R.~B.\ 1988, Nearby Galaxies
  Catalog (Cambridge: Cambridge University Press)
\bibitem[Vacca, Cushing, \& Rayner(2003)]{vacca03} Vacca, W.~D.,
  Cushing, M.~C., \& Rayner, J.~T. 2003, \pasp, 115 ,389
\bibitem[Van Dyk et al.(2000)]{vandyk00} Van Dyk, S.~D., et al.~2000,
  \pasp, 112, 1532
\bibitem[Wade \& Horne(1988)]{wade88} Wade, R.~A., \& Horne, K. 1988,
  \apj, 324, 411
\bibitem[Weiner et al.(2005)]{weiner05} Weiner, B.~J., et al. 2005,
  \apj, 620, 595
\bibitem[Wells et al.(1994)]{wells94} Wells, L.~A., et al. 1994, \aj,
  108, 2233
\bibitem[Yamaoka \& Itagaki(2009)]{yamaoka09} Yamaoka, H., \& Itagaki,
  K. 2009, CBET 1874, 1
\bibitem[Yoon \& Cantiello(2010)]{yoon10} Yoon, S.-C., \& Cantiello, M. 2010, \apj, 717, L62

\end{thebibliography}
\end{document}